\documentclass[twocolumn,showpacs,preprintnumbers,amsmath,amssymb]{revtex4}

\usepackage{graphicx}
\usepackage{dcolumn}
\usepackage{bm}

\begin{document}

\title{Role of low-$l$ component in deformed wave functions near the 
continuum threshold}

\author{Kenichi Yoshida$^{1}$}
\author{Kouichi Hagino$^{2}$}
\affiliation{
$^{1}$Department of Physics, Graduate School of Science, Kyoto University, 
Kyoto 606-8502, Japan
\\
$^{2}$Department of Physics, Graduate School of Science, 
Tohoku University, Sendai 980-8578, Japan}%

\date{\today}

\begin{abstract}
The structure of deformed single-particle wave functions 
in the vicinity of zero energy limit is 
studied using 
a schematic model with a quadrupole deformed finite square-well potential. 
For this purpose, we expand the single-particle wave functions in multipoles 
and seek for the bound state and the Gamow resonance solutions. 
We find that, for the $K^{\pi}=0^{+}$ states, 
where $K$ is the $z$-component of the orbital angular momentum, 
the probability of each multipole components in the deformed wave
function 
is connected  
between the negative energy and the positive energy regions
asymptotically, although it 
has a discontinuity around the threshold. 
This implies that the $K^{\pi}=0^{+}$ resonant level exists physically 
unless the $l=0$ component is inherently large when extrapolated to the well bound region. 
The dependence of the multipole components on deformation is also discussed. 
\end{abstract}

\pacs{21.10.Pc, 21.60.Jz, 24.30.Gd}
\maketitle

\section{Introduction}
Physics of nuclei located 
far from the $\beta$ stability line 
has been one of the main current subjects of nuclear physics. 
One of the unique properties of drip-line nuclei is that the Fermi 
level lies close to zero. 
Understanding of single-particle levels in the continuum is essential in 
describing the nuclear structure close to, and beyond, the drip line, since 
the shell structure of both bound and continuum levels 
plays an important role in many-body correlations such as deformation 
and pairing. 

It has been argued recently that, as the binding energy approaches zero, 
the $s$-wave component of a bound single-particle wave function 
behaves uniquely in a deformed 
potential, and plays a dominant role in Nilsson levels 
with $\Omega^{\pi}=1/2^{+}$\cite{mis97,ham04}. 
Naively, resonant levels can be considered as an extension of 
bound states into the positive energy regime. 
Therefore, if the $s$-wave component keeps dominant in the continuum, 
the level with $\Omega^{\pi}=1/2^{+}$ might not exist as a physical state. 
Notice that, for a Nilsson Hamiltonian\cite{nil55}, single-particle
levels with $\Omega=1/2$ belonging to high-$j$ orbit 
comes down in energy in a prolately deformed potential. 
These states play an important role in 
generating the deformed shell structure. 
It is therefore crucially important to investigate 
the role of low-$l$ component in a deformed wave function 
for $\Omega=1/2$ states and its transition from bound to resonant levels. 

The structure of deformed single-particle levels in the continuum 
has been investigated in a few publications. 
In Ref.\cite{fer97}, the resonance energy of negative parity states 
was studied by employing the Gamow wave function. 
The Analytic Continuation in the Coupling Constant (ACCC) method was applied to study 
single-particle resonance states in spherical and deformed nuclei~\cite{cat00}. 
Using the multi-channel scattering approach, 
Ref.~\cite{ham05} has studied how the single-particle 
energies change from bound to resonant levels 
when the depth of the potential is varied. 
In order to fully understand the structure of deformed 
single-particle levels in the continuum, however, 
a detailed study of the wave 
function components is still necessary, 
in addition to the resonance energy itself. 

In this paper, we investigate the structure of deformed wave functions 
around zero energy using the Gamow state representation for a resonant 
state. To this end, we use a schematic model: 
a $Y_{20}$ deformed finite square-well potential 
without spin-orbit force. This enables us to determine the single-particle 
wave function analytically. 
To use the Gamow state for resonance 
has a certain advantage in analyzing the deformed wave function. 
That is, we are able to treat the bound and the resonant levels on the same 
footing, because the Gamow states are normalizable 
just like the bound states \cite{rom68}. 
It is then straightforward to see how the fraction of each component in 
the deformed wave functions changes when the single-particle level changes 
its character from bound to resonant. 
A slight disadvantage of this approach is that the expectation value  
with the Gamow states, including the probability of wave function components,  
becomes complex numbers. 
However, this is not a big defect for our purpose, since 
the physical quantity of the expectation values can be obtained by taking 
their real part\cite{ber96,aoy98}. 

The paper is organized as follows. In the next section, we present our 
model for a deformed single-particle potential. 
Numerical results and discussion are given in Sec.\ref{sec:results}. 
Finally, we summarize the paper in Sec.\ref{sec:summary}.

\section{\label{model}Model}

Our purpose is to study the structure of wave function in a deformed 
single-particle potential. To this end, 
we employ a schematic model for the single-particle potential, that is, 
a deformed square-well potential without the spin-orbit force, 
\begin{equation}
V(\boldsymbol{r})=-V_{0}\,\theta(R(\hat{\boldsymbol{r}})-r), 
\end{equation}
where 
$R(\hat{\boldsymbol{r}})=R_{0}(1+\beta_{2} Y_{20}(\hat{\boldsymbol{r}}))$. 
For simplicity, 
we expand this potential up to the first order of deformation parameter 
$\beta_{2}$ and obtain 
\begin{equation}
V(\boldsymbol{r})
\simeq  -V_{0}\left[\theta(R_{0}-r)+
R_{0}\beta_{2} Y_{20}(\hat{\boldsymbol{r}}) \delta(r-R_{0})\right]. 
\end{equation}
In order to solve 
the Schr\"odinger equation with this potential, 
we expand the wave function in the multipoles as 
\begin{equation}
\Psi_{K}(\boldsymbol{r})=\sum_{l}
\dfrac{u_{lK}(r)}{r}Y_{lK}(\hat{\boldsymbol{r}}),
\end{equation}
where the quantum number $K( =\Lambda)$ is the 
$z$-component of the orbital angular momentum $l$. 
By projecting out each multipole component, 
we obtain the coupled equations for the radial wave functions given by 
\begin{multline}
\Bigl[ -\dfrac{\hbar^{2}}{2m}\dfrac{d^{2}}{dr^{2}}-V_{0}\theta(R_{0}-r)
+\dfrac{\hbar^{2}l(l+1)}{2mr^{2}}-E \Bigr]
u_{lK}(r) \\
=V_{0}R_{0}\beta_{2}\delta(r-R_{0})
\sum_{l^{\prime}}\langle lK|Y_{20}|l^{\prime}K 
\rangle u_{l^{\prime}K}(r). \label{eq:Sch2}
\end{multline}

For the positive energy solution, $E > 0$, we impose the 
boundary condition corresponding to the Gamow state for resonance. 
That is, the wave function is 
regular at the origin and satisfies the out-going boundary condition 
$u(r) \sim e^{ikr}$ asymptotically. 
This boundary condition is satisfied only if 
the energy is complex, 
$E=\hbar^{2}k^{2}/2m=E_{R}-i\Gamma/2$, 
where $E_{R}$ and $\Gamma$ are the resonance energy and the width, 
respectively. 
In the case for $\Gamma=0$ and $E_R<0$, the Gamow state wave function
is equivalent to 
the bound state wave function, which satisfies the decaying 
asymptotics $u(r) \sim e^{-\gamma r}$, where 
$\gamma=\sqrt{-2mE_R/\hbar^2}$. 
 
The solutions of the coupled-channels equations (\ref{eq:Sch2}) 
therefore read (we omit the subscript $K$ for simplicity of notation),
\begin{equation}
u_{l}(r)=
\begin{cases}
A_{l}\,rj_{l}(k_{1}r) &\hspace{0.2cm} (r < R_{0}), \\
B_{l}\,rh_{l}^{(+)}(kr) &\hspace{0.2cm} (r \geq R_{0}), 
\end{cases}
\end{equation}
where $k_{1}=\sqrt{2m(E+V_{0})/\hbar^{2}}, k=\sqrt{2mE/\hbar^{2}}$, 
and $j_{l}(x), h_{l}^{(+)}(x)$ are the spherical 
Bessel and Hankel functions, respectively. 
The amplitudes $A_{l}$ and $B_{l}$ are determined by the matching condition 
at $r=R_{0}$ given by,
\begin{align}
u_{l}(R_{-})=&u_{l}(R_{+}), \\
-\dfrac{\hbar^{2}}{2m}\left[ u_{l}^{\prime}(R_{+})-u_{l}^{\prime}(R_{-}) \right]=&
V_{0}R_{0}\beta_{2} \sum_{l^{\prime}} \langle lK|Y_{20}|l^{\prime}K \rangle u_{l^{\prime}}(R_{0}),
\end{align}
where $R_{\pm}$ represents $\lim_{\varepsilon \to 0}R_{0}\pm \varepsilon$. 

The bound state wave function is normalized as 
\begin{equation}
1=\int d\boldsymbol{r}\,|\Psi_K(\boldsymbol{r})|^2=
\sum_{l}N_{l}, 
\label{normtot}
\end{equation}
where 
\begin{equation}
N_{l}=\int_{0}^{\infty}\mathrm{d}r |u_{l}(r)|^{2}. 
\end{equation}
The Gamow state wave function can be also 
normalized by introducing the regularization 
factor as Zel'dovich proposed\cite{zel61} 
\begin{align}
N_{l}=&\lim_{\epsilon \to 0}
\int_{0}^{\infty}\mathrm{d}r e^{-\epsilon r^{2}}\{u_{l}(r)\}^{2} \\
=&\int_{0}^{R_{0}}\mathrm{d}r \{A_{l}\, rj_{l}(k_{1}r) \}^{2} \notag \\ 
&+ \lim_{\epsilon \to 0}
\int_{R_{0}}^{\infty}\mathrm{d}r e^{-\epsilon r^{2}}\{B_{l}\, rh_{l}^{(+)}(kr) \}^{2}.
\end{align}
Using a property of the spherical Bessel function\cite{mor53}, one 
can evaluate the first term as 
\begin{multline}
\int_{0}^{R_{0}}\mathrm{d}r\, \{A_{l}\,rj_{l}(k_{1}r) \}^{2}\\
=\dfrac{A_{l}^{2}R_{0}^{3}}{2}
\Bigl(\{j_{l}(k_{1}R_{0})\}^{2} - j_{l-1}(k_{1}R_{0})j_{l+1}(k_{1}R_{0}) 
\Bigr).
\end{multline}
The second term can be also evaluated 
using the contour integral method or equivalently the 
Complex Scaling Method (CSM). The 
result is given by \cite{gya70}, 
\begin{multline}
\lim_{\epsilon \to 0}\int_{R_{0}}^{\infty}\mathrm{d}r \,
e^{-\epsilon r^{2}}\{B_{l}\,r h_{l}^{(+)}(kr) \}^{2} \\
=-\dfrac{B_{l}^{2}R_{0}^{3}}{2}
\Bigl(\{h_{l}^{(+)}(kR_{0})\}^{2} - 
h_{l-1}^{(+)}(kR_{0})h_{l+1}^{(+)}(kR_{0}) \Bigr).
\label{normalization}
\end{multline}
Note that the fraction of multipole components 
$N_{l}$ is in general a complex number for the Gamow state wave function. 

\begin{figure}[tp]
\begin{center}
\includegraphics[scale=0.58]{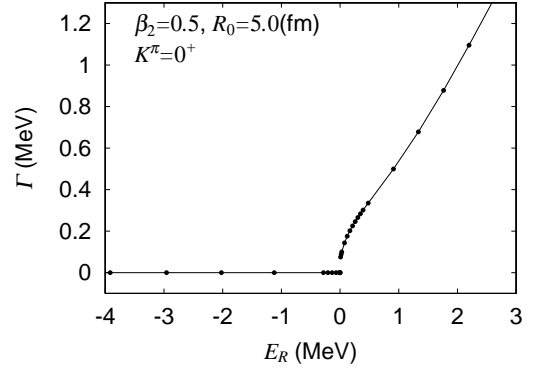}
\caption{The real part of the energy 
and the resonance width 
for a $K^{\pi}=0^{+}$ state with various potential depths. 
The corresponding potential depths are shown in Fig.\ref{Energy2}. 
}
\label{Energy}
\end{center}
\end{figure}

\section{\label{sec:results}Results and discussion}

Let us now discuss the behaviour of the low-$l$ components in deformed 
wave functions. 
In Sec. \ref{sec:V0}, we vary the potential depth for a fixed 
deformation parameter, while 
we vary the deformation parameter 
for a fixed potential depth in Sec. \ref{sec:def}.

\subsection{\label{sec:V0}Dependence on potential depth}

\begin{figure}[tp]
\begin{center}
\includegraphics[scale=0.58]{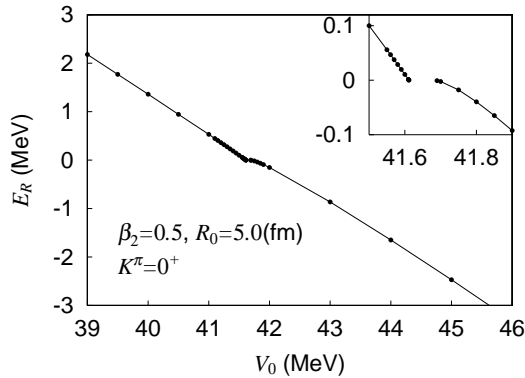}
\caption{The real part of the 
energy for $K^{\pi}=0^{+}$ state as a function of 
the potential depth.  
In the inset, the behaviour around zero energy is enlarged.}
\label{Energy2}
\end{center}
\end{figure}

\begin{figure}[tp]
\begin{center}
\includegraphics[scale=0.55]{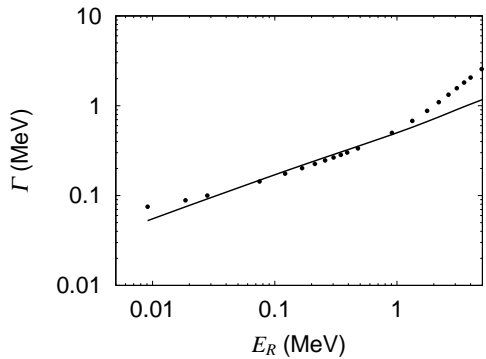}
\caption{
Same as Fig.\ref{Energy}, but  
in the logarithmic scale. 
The solid line is 
an expectation for the pure $s$-wave configuration given by 
Eq. (\ref{McVoy}).}
\label{Energy_fit}
\end{center}
\end{figure}

We first study the wave functions at a fixed deformation, $\beta_{2}=0.5$. 
Figure \ref{Energy} shows the real and 
imaginary parts of the energy for a $K^{\pi}=0^{+}$ state 
in varying the potential depth $V_0$. 
The correspondence between the potential depth and the real part of 
the energy is shown in Fig.\ref{Energy2}.
We observe that the width is quite large 
even for a small values of positive energy. 
This large width is caused by the admixture of the $l=0$ component 
in the wave function. 
Indeed, 
as shown in Fig.\ref{Energy_fit}, in the small positive energy 
region (0.1 MeV $< \Re(E) <$ 1.0 MeV), 
the behavior of the width is consistent with the relation 
expected for the $s$-wave resonance state \cite{voy67,fer97,ham05}, 
\begin{equation}
\Gamma \propto  \Re(E)^{l+1/2}\times \Re(N_{l})\Bigr|_{l=0}, 
\label{McVoy}
\end{equation}
where $\Re(E)$ denotes the real part of $E$. 

Below $E_{R}=$0.1 MeV, the width is larger than the solid line, 
which predicts $\Gamma =0$ at $E_R=0$.  
Also, we did not find a physical solution between $V_0$= 41.62 and 41.68 
MeV, as is shown in the inset of Fig.\ref{Energy2}. 
These might be related to the possible presence of the anti-bound and
`crazy' resonance states, as presented in Ref.~\cite{tan99} for a
spherical square-well potential (see Fig.1 of Ref.~\cite{tan99}). 
Above 1.0 MeV also, the width is larger than that expected by 
Eq.(\ref{McVoy}). 
This is due to the fact that the relation Eq.(\ref{McVoy}) is valid
only for small values of $k$~\cite{voy67}. 

In Fig.\ref{Energy2}, we see that the slope of the single-particle 
energy as a function of the potential depth, $\mathrm{d}E/\mathrm{d}V_{0}$,  
or equivalently $\mathrm{d}E/\mathrm{d}A$, where $A$ is the mass number, 
becomes smaller in approaching the zero binding energy. 
For a spherical square-well potential, it has been shown that  
$\mathrm{d}E_{l}/\mathrm{d}A \rightarrow 0$ for $l=0$ in the 
limit of zero binding~\cite{ham01}. 
This is due to the fact that the $s$-wave function can be easily extended 
outside the nuclear potential and also the kinetic energy is reduced 
due to the absence of the centrifugal barrier~\cite{ham01}. 
This property implies that the $l=0$ component becomes 
dominant in a deformed wave function around the zero-binding region. 
On the other hand, the slope has a finite value in the positive energy 
region even in the limit of zero energy, thus the slope has a discontinuity 
around zero energy. Therefore, a care must be taken, as discussed in Ref.~\cite{tan99}, 
when one estimates the energy of a deformed 
resonant level with $K^\pi=0^+$ by using the ACCC method~\cite{cat00}. 

\begin{figure}[tp]
\begin{center}
\begin{tabular}{cc}
\includegraphics[scale=0.7]{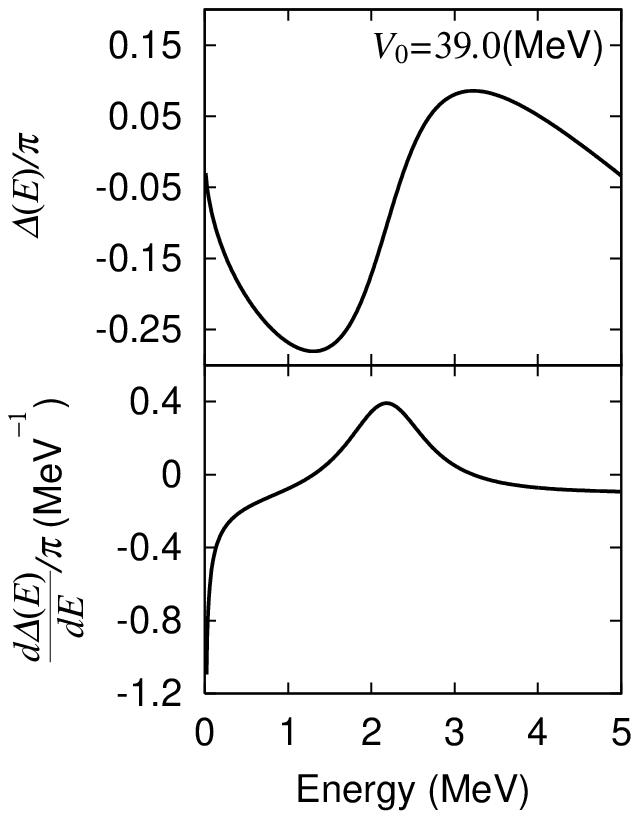}&
\includegraphics[scale=0.7]{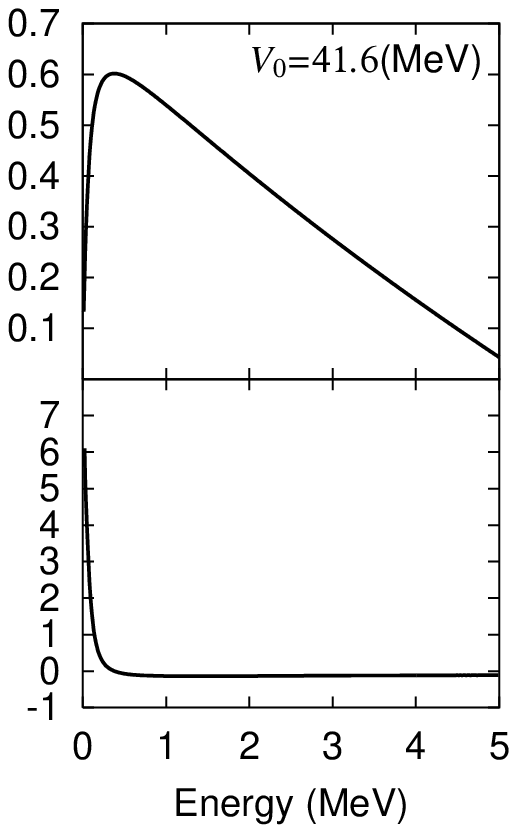}
\end{tabular}
\end{center}
\caption{The eigenphase sum and its energy derivative for 
$K^{\pi}=0^{+}$ state 
with potential depth $V_{0}=39.0$ MeV (the left panel), 
and $V_{0}=41.6$ MeV (the right panel).}
\label{eigenphase}
\end{figure}

The resonance energy and width can be also estimated using 
the eigenphase sum $\Delta(E)$ \cite{hag04}. 
It is defined in terms of the eigenvalues of the scattering matrix ($S$-matrix) as 
\begin{equation}
(U^{\dagger}SU)_{aa^{\prime}}=e^{2i\delta_{a}(E)}\delta_{a,a^{\prime}}, 
\hspace{0.2cm}
\Delta(E)=\sum_{a}\delta_{a}(E).
\end{equation}
The resonance energy and width are identified with the peak energy 
of $\mathrm{d}\Delta(E)/\mathrm{d}E$ 
and its FWHM, respectively\cite{mut01}. 
Figure \ref{eigenphase} shows the eigenphase sum for the $K^{\pi}=0^{+}$ state 
with two different potential depths. 
Comparing Figs.\ref{Energy}, \ref{Energy2} and \ref{eigenphase}, 
we see a good correspondence between the two definitions of resonance 
state, {\it i.e.,} the Gamow state representation and the 
approach with the eigenphase sum. 

\begin{figure}[tp]
\begin{center}
\includegraphics[scale=0.58]{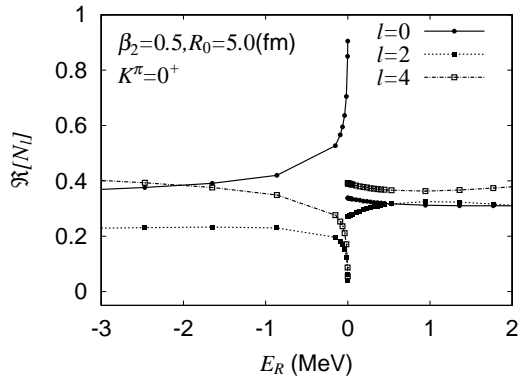}
\caption{The real part of the fraction for 
each multipole component $N_{l}$ for the $K^{\pi}=0^{+}$ state. 
The solid, dotted, and dot-dashed lines indicate the $l=0, 2$ and $l=4$ 
components, respectively.}
\label{real_component}
\end{center}
\end{figure}

\begin{figure}[tp]
\begin{center}
\includegraphics[scale=0.58]{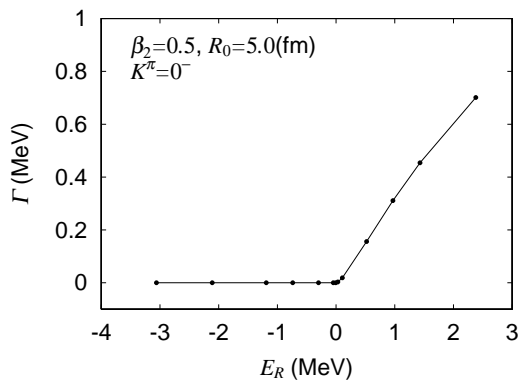}
\caption{Same as Fig.\ref{Energy}, but for a $K^{\pi}=0^{-}$ state.} 
\label{Energy_0-}
\end{center}
\end{figure}

We now discuss the energy dependence of the fraction of the multipole 
components in the deformed wave function. 
Figure \ref{real_component} shows the real part of the fraction for 
each multipole component 
in the Gamow state wave function with $K^{\pi}=0^{+}$. 
When the binding energy approaches zero, 
the $s$-wave component in the deformed wave function 
becomes dominant. In contrast, in the positive energy region, 
all the multipole components have a finite value even in the zero 
energy limit 
and show similarity with the well bound cases. 
As we will discuss in the next section (see 
Fig.\ref{Nilsson} below), the state shown in Fig.\ref{real_component} 
originates from the $2d$ orbit in the spherical limit. 
This states couples with the lower-lying $2s$, $1g$ and the 
higher-lying $3s$ states. 
The dominant component is $l=4$ both in well bound and in resonant levels, 
as one sees in Fig.\ref{real_component}. 
This suggests that both the well bound and the resonant levels have a 
similar property to each other and the intuitive picture that 
the resonant level is an extension of a bound level into the continuum 
is valid. 

Only at the limit of zero binding, 
the singular behavior of the $l=0$ component appears. 
This is entirely due to the property of the normalization integral, Eq.(\ref{normalization}). 
Since the Gamow state wave function is equivalent to the bound state wave function 
for $E_{R}<0, \Gamma=0$, Eq.(\ref{normalization}) holds both for the
resonance and the bound states. 
For small values of $k$, 
Eq.(\ref{normalization}) is proportional to
$k^{2l-1}$ as discussed in Refs.~\cite{mis97,RJM92}, 
that diverges only for $l=0$ as $k\to$ 0. 
When the total wave function $\Psi_K$ is normalized according to
Eq.(\ref{normtot}), then only the $s$-wave component is allowed in
the wave function \cite{mis97}. 
This condition is always met for the bound state when the binding
energy approaches the threshold. 
In principle, the same consideration can apply also
to the resonance state when the resonance energy approaches zero
from the positive energy side. 
However, as we show in
Fig.\ref{Energy}, the resonance state acquires a relatively large
width even when the real part of the energy is infinitesimally
small. Since $k$ is defined as $k=\sqrt{2m(E_{R}-i\Gamma/2)/\hbar^2}$, 
it remains a constant even if $E_R$ itself approaches zero. 
This leads to the disappearance of the ``$s$-wave dominance" in the
positive energy side. 

\begin{figure}[tp]
\begin{center}
\includegraphics[scale=0.58]{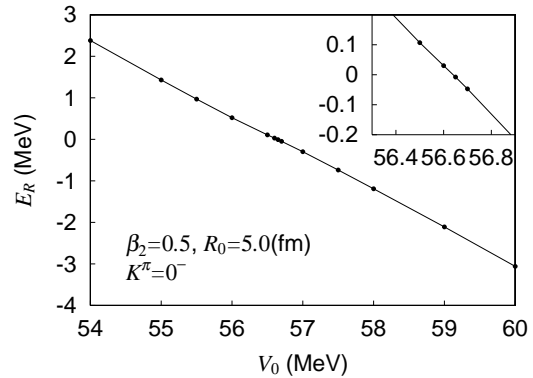}
\caption{Same as Fig.\ref{Energy2}, but for the $K^{\pi}=0^{-}$ state.}
\label{Energy2_0-}
\end{center}
\end{figure}

\begin{figure}[tp]
\begin{center}
\includegraphics[scale=0.58]{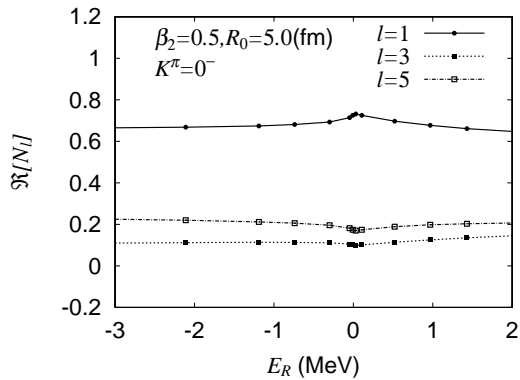}
\caption{The real part of the fraction for each multipole components 
$N_{l}$ for the $K^{\pi}=0^{-}$ state.
The solid, dotted, and dot-dashed lines indicate the $l=1, 3$ and $l=5$ 
components, respectively.
}
\label{real_component_0-}
\end{center}
\end{figure}

We next study the case for $K^{\pi}=0^{-}$. 
In Figs. \ref{Energy_0-} and \ref{Energy2_0-}, 
we show the dependence of the single-particle energy on the potential 
depth. 
In contrast to the case for $K^{\pi}=0^{+}$, 
due to the presence of the centrifugal barrier, 
we do not see 
any singular behavior around zero energy.  
Single-particle energies are connected smoothly when changing the 
potential depth, 
and the width increases gradually in the small positive energy region. 
Figure \ref{real_component_0-} shows the fraction of each multipole 
component in the Gamow state wave function. 
As the binding energy approaches zero, the 
$p$-wave component becomes relatively large, that is consistent with the 
dominance of low-$l$ component in the limit of zero binding energy 
discussed in Ref.\cite{ham04}. 
The fractions are connected smoothly and 
asymptotically in the bound and resonant regions. 

\subsection{\label{sec:def}Deformation dependence}
\begin{figure}[tp]
\begin{center}
\includegraphics[scale=0.58]{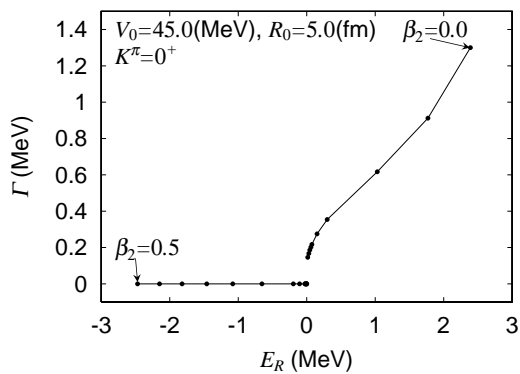}
\caption{Same as Fig.\ref{Energy} except for varying the deformation. 
This level corresponds to 
the one originating from the $2d$ orbit in the spherical limit. 
The deformation dependence of the single-particle energies 
is shown in Fig.\ref{Nilsson}. 
}
\label{Energy_V0=45}
\end{center}
\end{figure}

\begin{figure}[tp]
\begin{center}
\includegraphics[scale=0.58]{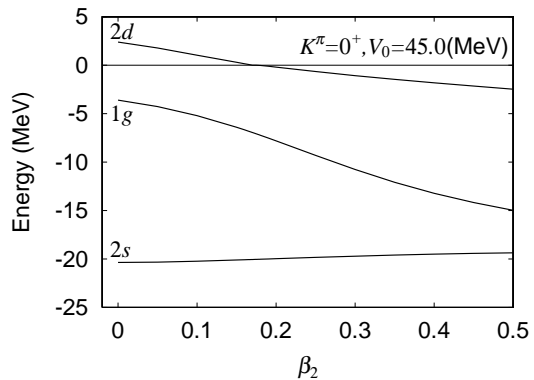}
\caption{Single-particle energies for the 
$K^{\pi}=0^{+}$ state as a function of deformation parameter $\beta_{2}$. 
The potential depth is $V_{0}=45.0$ (MeV), and 
the potential radius $R_{0}=5.0$ (fm).}
\label{Nilsson}
\end{center}
\end{figure}

In this subsection, 
we study the deformation dependence of the low-$l$ component 
in deformed wave functions for a fixed potential depth. 
In the realistic situation, the location of single-particle levels 
changes as a function of nuclear deformation. 
Especially, the levels of $\Omega=1/2$ ($K=0$) with (without) 
spin-orbit force belonging to high-$j$ (high-$l$) 
orbit in the spherical limit 
play an important role in nuclear deformation. 

Figure \ref{Energy_V0=45} shows the resonance energy and width 
when the deformation parameter is varied from 
$\beta_{2}=0.0$ to 0.5. 
The potential depth $V_0$ and the radius $R_0$ 
are set to be 45.0 MeV and 5.0 fm, respectively. 
This state belongs to the $2d$ orbit at 
$\beta_{2}=0.0$ as shown in Fig.\ref{Nilsson}. 
At around zero energy, we see the similar behavior as in Fig.\ref{Energy}: 
the width is quite large even for the small values of positive energy, 
which implies that the $l=0$ component 
is responsible for the width of the resonant level.

\begin{figure}[tp]
\begin{center}
\includegraphics[scale=0.58]{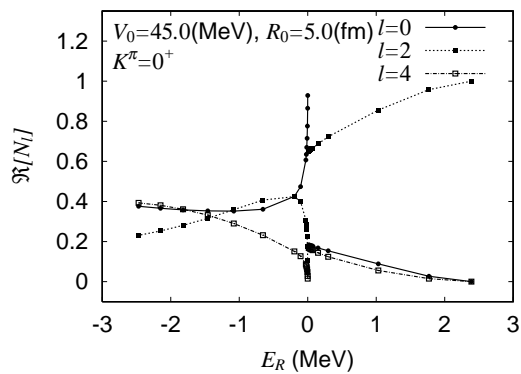}
\caption{Same as Fig.\ref{real_component} except for varying the deformation.}
\label{V0=45_component}
\end{center}
\end{figure}

The corresponding wave function components for this state are shown 
in Fig.\ref{V0=45_component}. 
As in Fig.\ref{real_component}, 
we see the singular behavior for the 
$s$-wave component at around zero-binding energy, 
corresponding to the ``$s$-wave dominance" in the limit of zero binding.  
Except for the zero-energy region, however, we see that the fraction 
of each multipole components is linked 
asymptotically 
and is smoothly connected to the $d$-state resonant level 
in the spherical limit. 
From this calculation, it is evident that the singular behavior of the 
$l=0$ component for the $K^{\pi}=0^{+}$ state 
occurs only just below the continuum threshold and 
this state is connected to the physical resonant level in the continuum. 
Furthermore, the fraction of each-$l$ components in the deformed wave function 
is connected smoothly from the bound to the 
resonant levels except for the region near the threshold. 

\section{\label{sec:summary}Summary}

We have analyzed the structure of the 
deformed wave functions around zero energy 
using the Gamow state wave function for resonance, with which  
one can treat the resonant and bound levels on the same footing and 
thus analyze the wave function continuously from the negative to the 
positive energy regions. 
For this purpose, we developed a 
schematic model with a deformed square-well potential. 
Since 
the wave functions can be obtained analytically with this model, 
detailed analyses of the deformed wave functions were possible. 
For a $K^{\pi}=0^{+}$ state, 
we have found a singularity in the resonance width as well 
as in the $s$-wave component 
in the deformed wave function at around zero energy. 
That is, 
the width becomes considerably large even in the 
small positive energy region and 
the $l=0$ component approaches unity in the limit of zero binding. 
We have shown that the ``$s-$wave dominance" 
occurs only at the threshold of continuum. 
Far from the zero energy region, the probability of each-$l$ components 
is connected asymptotically. 
This implies that the $K^{\pi}=0^{+}$ resonant level exists 
unless the $l=0$ component is large inherently when extrapolated to 
the well bound region. 
In contrast, for the $K^{\pi}=0^{-}$ state,
 we did not find any singular behavior even in the zero-energy limit. 
The single-particle energies are connected smoothly when changing the 
potential depth, 
and the width increases gradually in the small positive energy region.  
The probability of each-$l$ component in the wave function 
is also connected smoothly and asymptotically between the bound and the 
resonant regions. 

\begin{acknowledgments}
We thank Prof.~K.~Matsuyanagi for his continuing encouragement and 
for many useful discussions. We also thank Dr.~T.~Myo for useful discussions on the 
imaginary part of expectation values for a Gamow state. 
We acknowledge discussions with the member of the Japan-U.S. 
Cooperative Science Program
``Mean-Field Approach to Collective Excitations in Unstable Medium-Mass 
and Heavy Nuclei". 
This work was supported by the Japanese Ministry of Education, 
Culture, Sports, Science and Technology by Grant-in-Aid for Scientific 
Research under Contract number 16740129. 

\end{acknowledgments}

\end{document}